\documentclass[12pt]{article}
\usepackage{epsfig}
\usepackage{graphicx}
\usepackage{amsmath}
\usepackage{hhline}
\usepackage{amssymb}
\usepackage{times}
\usepackage{cite}

\newlength{\dinwidth}
\newlength{\dinmargin}
\begin{document}
\newcommand{\pom}{{I\!\!P}}
\newcommand{\reg}{{I\!\!R}}
\newcommand{\slowpi}{\pi_{\mathit{slow}}}
\newcommand{\fiidiii}{F_2^{D(3)}}
\newcommand{\fiidiiiarg}{\fiidiii\,(\beta,\,Q^2,\,x)}
\newcommand{\n}{1.19\pm 0.06 (stat.) \pm0.07 (syst.)}
\newcommand{\nz}{1.30\pm 0.08 (stat.)^{+0.08}_{-0.14} (syst.)}
\newcommand{\fiidiiiful}{F_2^{D(4)}\,(\beta,\,Q^2,\,x,\,t)}
\newcommand{\fiipom}{\tilde F_2^D}
\newcommand{\ALPHA}{1.10\pm0.03 (stat.) \pm0.04 (syst.)}
\newcommand{\ALPHAZ}{1.15\pm0.04 (stat.)^{+0.04}_{-0.07} (syst.)}
\newcommand{\fiipomarg}{\fiipom\,(\beta,\,Q^2)}
\newcommand{\pomflux}{f_{\pom / p}}
\newcommand{\nxpom}{1.19\pm 0.06 (stat.) \pm0.07 (syst.)}
\newcommand {\gapprox}
   {\raisebox{-0.7ex}{$\stackrel {\textstyle>}{\sim}$}}
\newcommand {\lapprox}
   {\raisebox{-0.7ex}{$\stackrel {\textstyle<}{\sim}$}}
\def\gsim{\,\lower.25ex\hbox{$\scriptstyle\sim$}\kern-1.30ex%
\raise 0.55ex\hbox{$\scriptstyle >$}\,}
\def\lsim{\,\lower.25ex\hbox{$\scriptstyle\sim$}\kern-1.30ex%
\raise 0.55ex\hbox{$\scriptstyle <$}\,}
\newcommand{\pomfluxarg}{f_{\pom / p}\,(x_\pom)}
\newcommand{\dsf}{\mbox{$F_2^{D(3)}$}}
\newcommand{\dsfva}{\mbox{$F_2^{D(3)}(\beta,Q^2,x_{I\!\!P})$}}
\newcommand{\dsfvb}{\mbox{$F_2^{D(3)}(\beta,Q^2,x)$}}
\newcommand{\dsfpom}{$F_2^{I\!\!P}$}
\newcommand{\gap}{\stackrel{>}{\sim}}
\newcommand{\lap}{\stackrel{<}{\sim}}
\newcommand{\fem}{$F_2^{em}$}
\newcommand{\tsnmp}{$\tilde{\sigma}_{NC}(e^{\mp})$}
\newcommand{\tsnm}{$\tilde{\sigma}_{NC}(e^-)$}
\newcommand{\tsnp}{$\tilde{\sigma}_{NC}(e^+)$}
\newcommand{\st}{$\star$}
\newcommand{\sst}{$\star \star$}
\newcommand{\ssst}{$\star \star \star$}
\newcommand{\sssst}{$\star \star \star \star$}
\newcommand{\tw}{\theta_W}
\newcommand{\sw}{\sin{\theta_W}}
\newcommand{\cw}{\cos{\theta_W}}
\newcommand{\sww}{\sin^2{\theta_W}}
\newcommand{\cww}{\cos^2{\theta_W}}
\newcommand{\trm}{m_{\perp}}
\newcommand{\trp}{p_{\perp}}
\newcommand{\trmm}{m_{\perp}^2}
\newcommand{\trpp}{p_{\perp}^2}
\newcommand{\alp}{\alpha_s}

\newcommand{\alps}{\alpha_s}
\newcommand{\sqrts}{$\sqrt{s}$}
\newcommand{\LO}{$O(\alpha_s^0)$}
\newcommand{\Oa}{$O(\alpha_s)$}
\newcommand{\Oaa}{$O(\alpha_s^2)$}
\newcommand{\PT}{p_{\perp}}
\newcommand{\JPSI}{J/\psi}
\newcommand{\sh}{\hat{s}}
\newcommand{\uh}{\hat{u}}
\newcommand{\MP}{m_{J/\psi}}
\newcommand{\PO}{I\!\!P}
\newcommand{\xbj}{x}
\newcommand{\xpom}{x_{\PO}}
\newcommand{\ttbs}{\char'134}
\newcommand{\xpomlo}{3\times10^{-4}}
\newcommand{\xpomup}{0.05}
\newcommand{\dgr}{^\circ}
\newcommand{\pbarnt}{\,\mbox{{\rm pb$^{-1}$}}}
\newcommand{\gev}{\,\mbox{GeV}}
\newcommand{\WBoson}{\mbox{$W$}}
\newcommand{\fbarn}{\,\mbox{{\rm fb}}}
\newcommand{\fbarnt}{\,\mbox{{\rm fb$^{-1}$}}}
%
%
\newcommand{\qsq}{\ensuremath{Q^2} }
\newcommand{\gevsq}{\ensuremath{\mathrm{GeV}^2} }
\newcommand{\et}{\ensuremath{E_t^*} }
\newcommand{\rap}{\ensuremath{\eta^*} }
\newcommand{\gp}{\ensuremath{\gamma^*}p }
\newcommand{\dsiget}{\ensuremath{{\rm d}\sigma_{ep}/{\rm d}E_t^*} }
\newcommand{\dsigrap}{\ensuremath{{\rm d}\sigma_{ep}/{\rm d}\eta^*} }
\newcommand{\dedx}{\ensuremath{{\rm d} E/{\rm d} x}}
\def\Journal#1#2#3#4{{#1} {\bf #2} (#3) #4}
\def\NCA{Nuovo Cimento}
\def\RPP{Rep. Prog. Phys.}
\def\ARNPS{Ann. Rev. Nucl. Part. Sci.}
\def\NIM{Nucl. Instrum. Methods}
\def\NIMA{{Nucl. Instrum. Methods} {\bf A}}
\def\NPB{{Nucl. Phys.}   {\bf B}}
\def\NPPS{Nucl. Phys. Proc. Suppl.}
\def\NPPSC{{Nucl. Phys. Proc. Suppl.} {\bf C}}
\def\PR{Phys. Rev.}
\def\PLB{{Phys. Lett.}   {\bf B}}
\def\PRL{Phys. Rev. Lett.}
\def\PRD{{Phys. Rev.}    {\bf D}}
\def\PRC{{Phys. Rev.}    {\bf C}}
\def\ZPC{{Z. Phys.}      {\bf C}}
\def\EJC{{Eur. Phys. J.} {\bf C}}
\def\EPL{{Eur. Phys. Lett.} {\bf}}
\def\CPC{Comp. Phys. Commun.}
\def\NP{{Nucl. Phys.}}
\def\JPG{{J. Phys.} {\bf G}}
\def\EPC{{Eur. Phys. J.} {\bf C}}
\def\PRSL{{Proc. Roy. Soc.}} {\bf}
\def\PETF{{Pi'sma. Eksp. Teor. Fiz.}} {\bf}
\def\JETPL{{JETP Lett}}{\bf}
\def\IJTP{Int. J. Theor. Phys.}
\def\HJ{Hadronic J.}




\begin{flushleft}
{\tt \today } \\
\end{flushleft}
\begin{center}
\begin{Large}
{\boldmath \bf Man made global warming explained - closing the blinds } \\

\end{Large}

\begin{flushleft}


T. Sloan. \\
(Dept of Physics, University of Lancaster)\\
A.W. Wolfendale, \\
(Dept. of Physics, University of Durham) \\

\end{flushleft}
\end{center}


\begin{abstract}
\noindent

   One of the big problems of the age concerns 'Global Warming', and whether
it is 'man-made' or 'natural'. Most climatologists believe that it is very 
likely to be the former but some scientists (mostly non-climatologists) 
subscribe to the latter. Unsurprisingly, the population at large is often 
confused and and is not convinced either way. Here we try to explain the 
principles of man-made global warming in a simple way. Our purpose is to 
try to understand the story which the climatologists are telling us through 
their rather complicated general circulation models. By limiting our 
attention to latitudes in the Southern Hemisphere we minimise 
the effects of industrial and land-generated aerosols. Here we only 
consider carbon dioxide and methane  and their effect on water vapour. 
The simple model comprising mainly the direct heating from the absorption of infrared 
radiation, illustrates the main principles of the science involved. The 
predicted temperature increase due to the increase of greenhouse gases in 
the atmosphere over the last century follows roughly the observed 
temperature increase. 

\end{abstract}






\section{The simple model}

The climate is very complex with many simultaneously changing
phenomena. This complexity serves to confuse both scientist and
layman alike. To illustrate the physics behind the global warming
caused by greenhouse gases, we describe a simple calculation in
which the complexity is neglected. Nothing that we say is new, but by
concentrating on fundamentals we hope to bring out the basic physics.
The calculation assumes that the Earth is warmed by the sun, reaching
an equilibrium temperature, $T$, at which the energy re-radiated
into space is balanced by the energy absorbed. We assume
that the atmosphere equalises the temperature so that the
absolute temperature in degrees Kelvin
is roughly uniform over the globe. Under these conditions
the energy radiated, $E$ watts per m$^2$, from the Earth system
follows Stefan's Law
\begin{equation}
E=k T^4
\label{eq1}
\end{equation}
where $k$ is a constant number. Here the 'Earth system' means the Earth
and its atmosphere.
If the total energy absorbed by the Earth system changes by an amount
$\Delta E$, it follows from equation \ref{eq1} that the change in
temperature to re-reach equilibrium will be given by
\begin{equation}
\frac{\Delta T}{T}=\frac{1}{4} \frac{\Delta E}{E}.
\label{eq2}
\end{equation}

At the temperature of the Earth the re-radiated
energy is in the infrared region of the spectrum. In this region
there is much absorption of energy by the so called greenhouse
gases: water vapour,
carbon dioxide (CO$_2$), methane (CH$_4$), ozone (O$_3$)
and other impurities
present in the atmosphere. Note that the main constituents of
the atmosphere (oxygen and nitrogen) do not absorb infrared
radiation since they are symmetric molecules with zero electric
dipole moments.  Without the greenhouse gases the average temperature of
the Earth (from equation \ref{eq1}) would settle to 255K (-18$^\circ$C).
This is too low for life as we know it to exist since liquid water would
be scarce at such temperatures.
However, the energy absorbed by the small concentrations
of greenhouse gases allows the atmosphere to act as a blanket for
the Earth, warming it to a more comfortable average of 14$^\circ$C.
Man-made increases of the greenhouse gases are then thought to
produce further warming on top of this i.e. man-made global warming.

In this note we describe calculations using the simple model to derive
the increase in temperature due to the increase in the greenhouse gases
CO$_2$ and CH$_4$ in the atmosphere observed over the last century. 
The response of water vapour (WV) to the increased temperature is 
also discussed. The absorption and re-emission of energy radiated 
from the Earth and its atmosphere, at a fixed temperature, is
computed in the program MODTRAN \cite{MODTRAN}.
The program is used to compute the radiated energy from the Earth 
system for a given set of concentrations of CO$_2$ or CH$_4$
allowing for all known absorption and re-emission effects of
radiation in the atmosphere.
The resulting change in the mean temperature of
the Earth can then be derived from the change in energy radiated as the
greenhouse gas concentration is changed, using equation \ref{eq2}.
The MODTRAN programme, though complex, is straightforward and simulates
the absorption, re-emission and scattering of the infra-red radiation in
the atmosphere. It should not be
confused with the more complicated climatological models which give somewhat
disparate results although all show an upward trend of temperature with
increased CO$_2$ and methane. These
differences are used by some 'man-made global warming skeptics'
as reasons for refusing
to accept the overall man-made explanation. They have no relevance, here.

Similar calculations to ours were published previously by Bellamy and
Barrett \cite{BandB}.

\section{The heating effects of greenhouse gases}

Figure \ref{fig1} \cite{PandO} illustrates the processes involved. 
The left hand upper curve (fig. \ref{fig1}a) shows the input radiation
from the sun (left hand curve) and the outgoing radiation from the Earth
(right hand curve) each plotted against the wavelength of the radiation.
The latter is the curve for the Earth as a perfect radiator without
the effects of greenhouse gases.
Under these conditions its mean temperature would reach an equilibrium
value of 255K (-18C) when the energy re-radiated balances that coming
from the sun.

The lower curves (figure \ref{fig1}d) show how the radiation is
absorbed by the greenhouse gases in the atmosphere (and fig \ref{fig1}b 
gives the total absorbance).  
The graphs show the fraction of the radiation which is
absorbed by each gas in the atmosphere plotted against the wavelength
of the radiation. With such absorption of heat the atmosphere acts like
a blanket allowing the Earth system to reach a new
more comfortable average equilibrium temperature of 287K(13.7$^\circ$C).


The addition of extra CO$_2$ and CH$_4$ will cause a further increase
in temperature. In the case of CO$_2$ this is caused by the extra
absorption in the wings of the band between wavelengths 13 and 18
microns (see figure \ref{fig1}).  This makes the band appear wider
so that the transparent gap between wavelengths of 8 to 13 microns
becomes narrower i.e. the blinds referred to in the title are being
closed.  The second and third rows of Table 1 shows the changes in the
concentrations of these greenhouse gases since industrialisation
started in about 1850 \cite{IPCC}.
The fourth row shows the energy absorbed in the
increased greenhouse gas concentration, as computed in MODTRAN
applying our simple model.
The last row shows the necessary increase in the temperature of
the Earth system to re-establish equilibrium with the amount of energy
re-radiated balancing that falling on the Earth  (computed according
to equation \ref{eq2}).

Table 1

\noindent
Variation of greenhouse gas concentrations (in parts per million, ppm)
with time and the calculated change in energy and temperature using
our simple model.
\vspace{-3mm}
\begin{center}
\begin{tabular}{|c|c|c|c|c|c|c|c|}\hline
 Date & 1850 & 1875 & 1900 & 1925 & 1950 & 1975 & 2000 \\
\hline
CO$_2$ Concentration (ppm) & 286 & 289 &  297 & 304 & 310 & 331 & 369 \\
CH$_4$ Concentration (ppm) &0.79 &0.82 & 0.86 &0.95 & 1.05&1.29 &1.56 \\
Re-radiated energy decrease (W/m$^2)$ & & & & & & & \\
at fixed Earth system temperature. 
& 0.0 &0.09 &0.25 & 0.41&0.60 &1.04 &1.70\\
Temperature increase ($^\circ$C) & 0.0 & 0.03 &0.08 &0.12 &0.18 &0.31 &0.51\\
needed to maintain equilibrium & & & & & & & \\  
\hline
\end{tabular}
\end{center}

The energy radiated to space from the Earth takes place from layers
of the upper troposphere below the stratosphere.  Here the atmosphere
is thin enough not to absorb much of the energy radiated from below. 
As the greenhouse gas concentration increases this altitude moves to 
a higher level where the temperature would normally 
be lower. However to radiate the extra energy equation \ref{eq1}  
shows that the temperature at this level must increase 
(final row of Table 1) to re-establish the equilibrium.
Below this altitude the heat in the Earth's atmosphere circulates 
mainly by 
convection, a process that is understood. The temperature in the
atmosphere decreases linearly from that at ground level 
to roughly -55$^\circ$C in the stratosphere.
This is easily understood from the thermodynamics of the atmosphere
\cite{Fred} which predict that the rate of fall of temperature
with altitude (the
so called lapse rate) is fixed. Assuming such a fixed lapse rate,
the changes in
temperature given in the last row of table 1 are transmitted to the
Earth's surface.

Figure \ref{fig2} shows the measured mean surface
temperature of the Earth as a function of time since 1880 
in a region chosen where industrial aerosols are almost absent (24-90 
$^\circ$S). 
Some of the variations on these data (at the level of 0.1$^\circ$C)
can be explained in terms of
large volcanic eruptions (the major dips, such as that due to Agung in 1963 
and Pinatubo in 1991), ozone variations, El Nino events and
the 'Southern Oscillation' (oceanic changes) \cite{AWW}. 
The curves shows the predicted temperature rises
using the simple model described above (the solid curve is the 
last row in table 1).

It can be seen that the calculations, based on these simple
physical principles, ignoring all complications, give a reasonable
explanation of the measurements to date with values which are not far
from those observed. 

\section{Discussion of the Results}

\begin{figure}[ht]
\includegraphics[width=33pc]{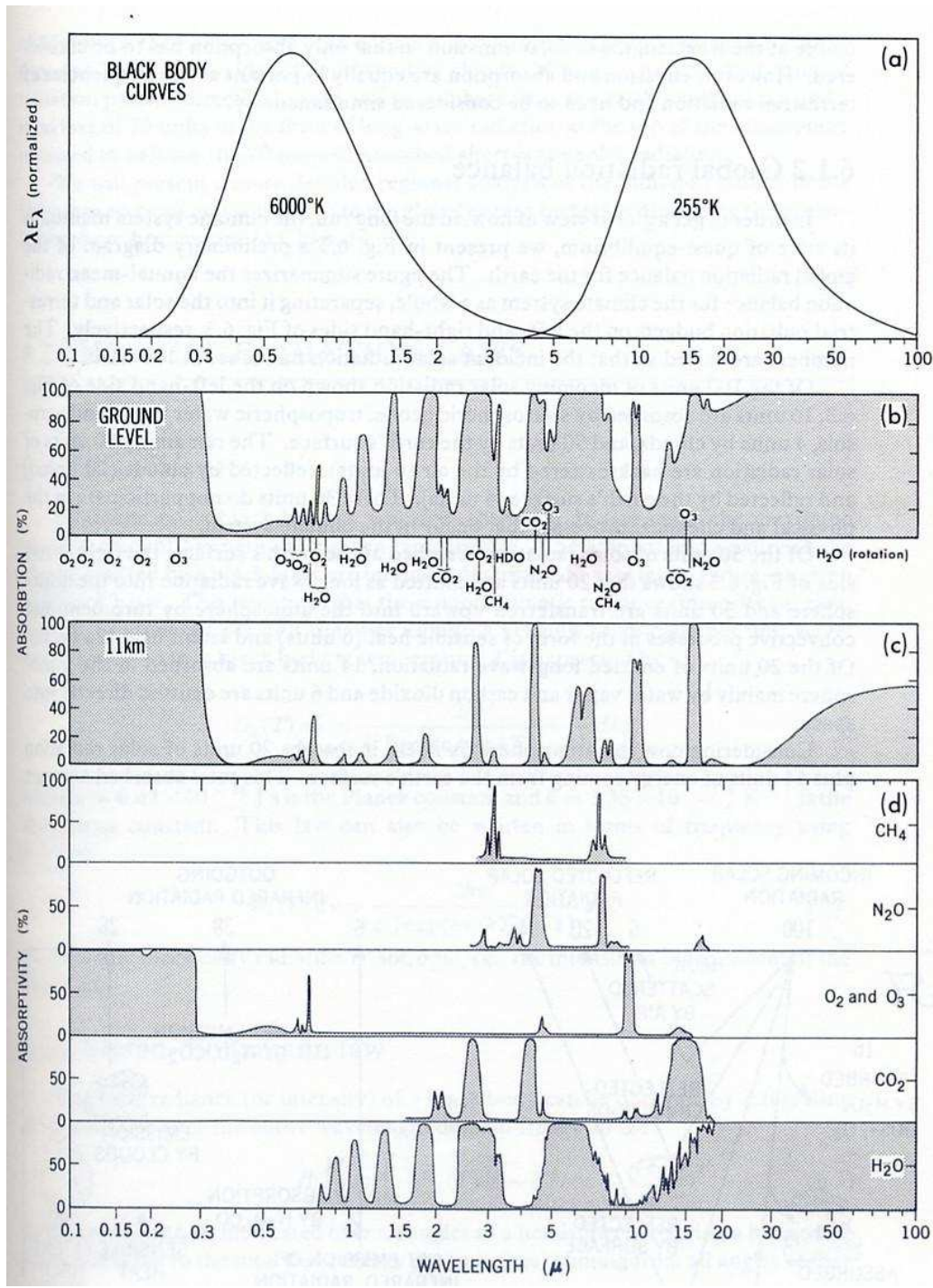}
\caption{\label{fig1} (a) The left hand upper curve shows the
intensity of the radiant energy reaching the Earth from the sun
which is mainly in the visible part of the spectrum (0.4 to 0.8 microns
wavelength). The right hand upper curve shows the intensity of the
radiation emitted by the Earth as a perfect radiator with no absorption
by greenhouse gases. This is at a much longer wavelength in the infrared 
part of the spectrum. 
(b) shows the total absorbance for the entire vertical extent of the 
atmosphere and (c) for the portion of the atmsophere above an altitude 
of 11 km. (d) show the fractions of the radiation at each wavelength 
absorbed in the atmosphere by each major greenhouse gas.  
Note that most of the energy re-radiated
by the Earth is absorbed in the atmosphere except in the gap from
8-13 microns wavelength.}
\end{figure}

\begin{figure}[ht]
\includegraphics[width=40pc]{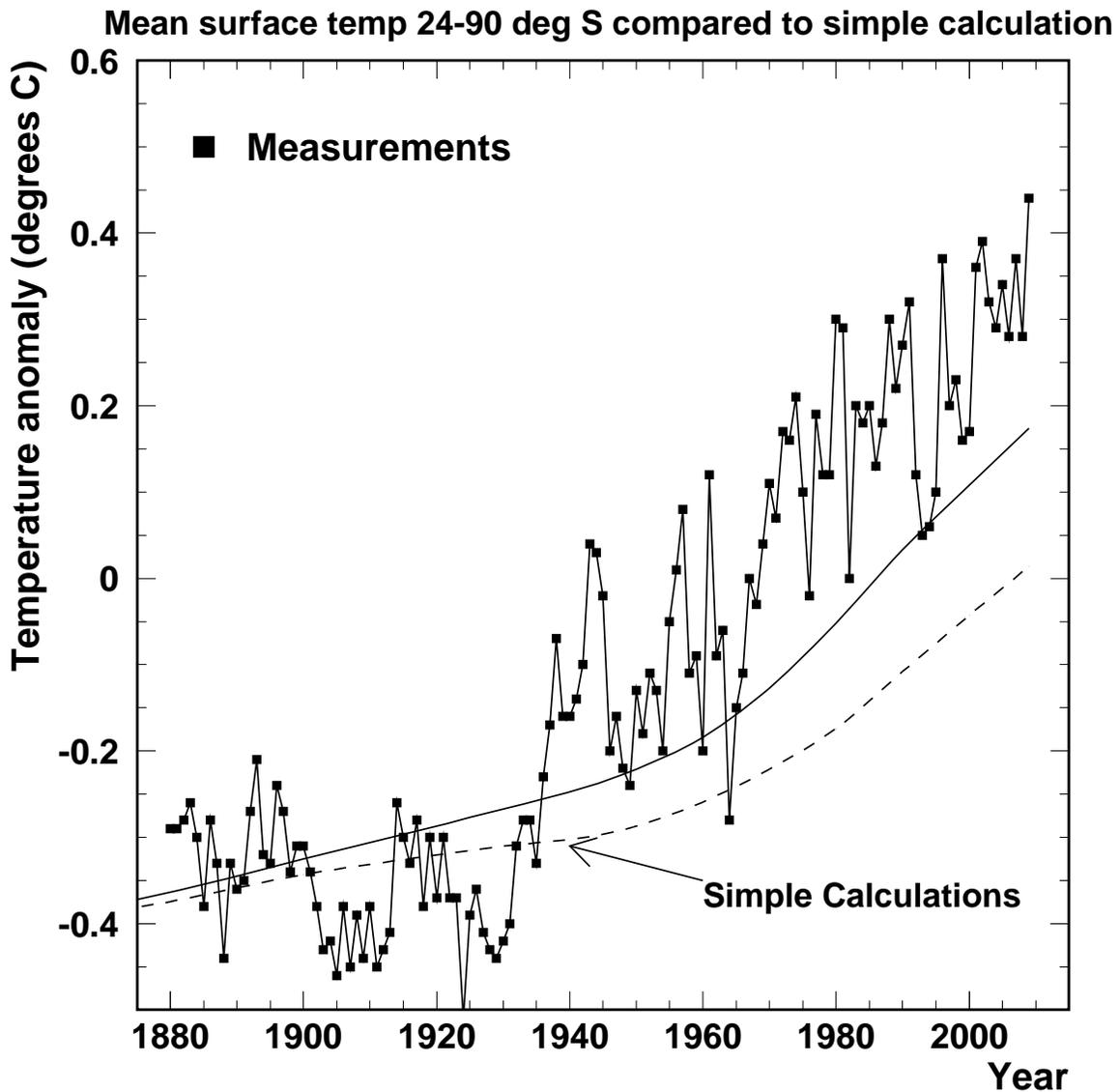}
\caption{\label{fig2} The points joined by the solid lines
show the measurements of the average annual
mean surface temperature of the Earth from meteorological stations
as a function of time since
1880 \cite{nasa} in the Southern Hemisphere 
where industrial aerosols are almost absent. The 
dashed curve  
shows the change in temperature predicted by the simple model where 
the CO$_2$ concentration only increases from its 1850 
value. The 
solid curve shows the behaviour when the CO$_2$ and CH$_4$ 
are changed together since 1850.  
}
\end{figure}

The simple calculation based on infrared absorption roughly reproduces
the observations, demonstrating the underlying physical principles of
the more complicated climate models. However,
the calculations shown in figure \ref{fig2} seem to
increase somewhat more slowly than the measurements.  
Furthermore, these calculations overestimate the overall warming since 
some of the energy is absorbed by the oceans rather than radiated away. 
The temperature rise in the oceans is much slower than in the atmosphere 
because of their large mass (this will eventually lead to delayed 
future warming).  
This shows that the absorption of infra-red
radiation by the atmosphere is not the only contributing process.
There are also other processes which we have ignored such as
the increases due to other greenhouse gases
(eg NOx, CFCs etc) as well as all other complications. 
We have also ignored feedback effects. One positive feedback is that the 
warming increases the amount of water vapour in the atmosphere by 
evaporation. Water 
is a good greenhouse gas (see figure \ref{fig1}), so warming causes 
more infra red radiation to be absorbed which in turn produces further 
warming, hence the term positive feedback. Such extra absorption is in 
the wavelength range 6-10 microns (see fig \ref{fig1}) closing the gap 
from the left (whereas CO$_2$ closes the gap from the right).   
There are also 'negative' (i.e. cooling) feedbacks such as those from 
atmospheric aerosols which reflect away sunlight so that their 
increase by industry causes cooling.  
In addition there is a negative feedback from clouds. 

\begin{figure}[ht]
\includegraphics[width=30pc]{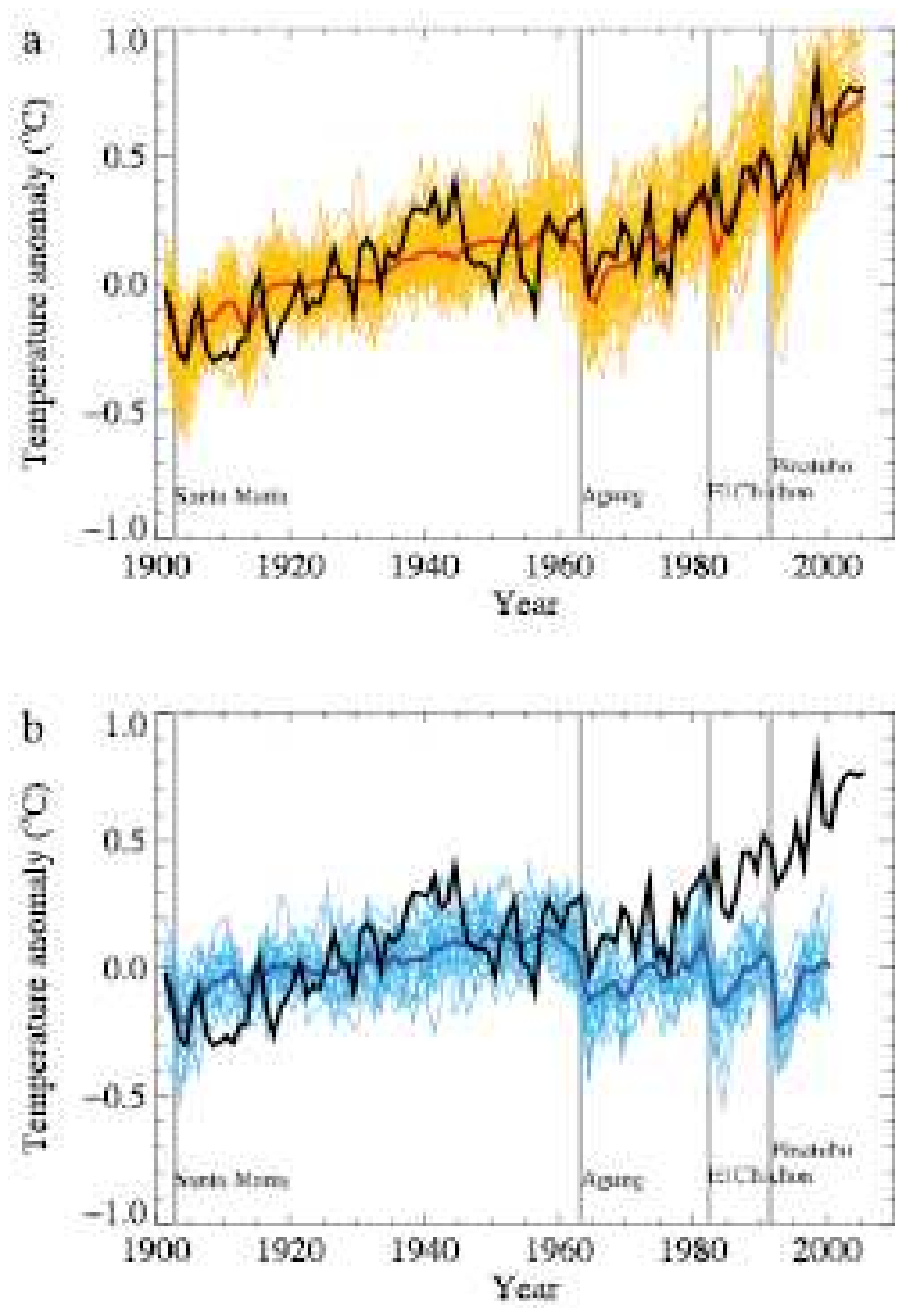}
\caption{\label{fig3} a) - Global mean surface temperatures
over the 20th century from observations (black) and as obtained
from 58 simulations produced by 14 different models driven by
both natural and human-caused factors that influence climate (yellow)
(red is the mean). b) shows 19 simulations from 5  models
with natural forcings only.
Temperature anomalies are shown relative to the
1901 to 1950 mean. Vertical grey lines indicate the timing of major
volcanic eruptions. This is Figure 9.5 from the IPCC Fourth Assessment
Report WG1 report (2007).   }
\end{figure}

The Intergovernmental Panel on Climate Change (IPCC), using their more
exact models for the whole Globe, produce the predictions shown in
figure \ref{fig3}. It can be seen that the warming of the last century 
(black curve in figure \ref{fig3}) 
is well reproduced by the models only if the man-made greenhouse gases
are included, although it will be noted that the excess in the region of 1940 found in our limited Southern region is also present here as is `our' deficit near 1910. The 1940 excess could possibly be ascribed to the effect of an 
El Nino event. 
The IPCC estimates that doubling 
the CO$_2$ level in the atmosphere will change the mean surface
temperature of the Earth by between 2 and 3.5$^\circ$C. Our
model predicts a rise of $\sim$1.3$^\circ$C (see also
\cite{BandB} who ignore all effects except infra red absorption),
again illustrating that something extra is needed
beyond the simple absorption of the infra-red radiation.

Most climatologists subscribe to the view that the global
warming over the last century is man made. Nevertheless, the IPCC are
not completely certain and only say that it is ``very likely'' that the
global warming since industrialization is man-made. Why is the IPCC
not completely certain? A serious problem is `how to deal with clouds?'.  These entities are highly variable in space and time and are complex.  It is also
possible that the climatologists may have made a
mistake and one or more of the many other contributing processes,
alluded to above, have margins of error which are bigger than
currently thought. This leads to the only argument that might be invoked to justify `doing nothing' about man-made emissions; this is to assume that the inevitable warming from anthropogenic gases is nullified by a process such as from man-made aerosols (a not uncommon claim some decades ago).  The undoubted warming is then due to unknown natural causes.  There are three objections to this scenario:

\begin{itemize}
\item[(a)] The ranks of climatologists must have made serious errors. We know of no such errors, although, undoubtedly small 
changes will result from future measurments and analyses.. 
\item[(b)] Our analysis of the man-made aerosol-poor region in figure \ref{fig2} (the Southern hemisphere)  mitigates against the problem of aerosols.
\item[(c)] In our so-far unsuccessful attempt to find a suitable natural cause,  none of the following have proved acceptable: meteoric dust, changes in the frequency of volcanoes, oceanic temperature re-distributions, geothermal emission changes.  Furthermore, no changes in the obliquity of the Earth's axis or Earth-sun distance, nor of solar irradiance of sufficient magnitude have occurred in the last Century (unlike in history when major temperature variations followed such changes).
    
\end{itemize}

In the face of this it is prudent to do something now if only as an
insurance policy. Otherwise we would be relying on an unexpected
future cancellation due to a completely unknown mechanism to save us from the possible ravages of climate
change induced by ever growing amounts of greenhouse gases in the
atmosphere.



\newpage

\end{document}